\pdfoutput=1

\documentclass[11pt]{article}

\usepackage[preprint]{acl}

\usepackage{times}
\usepackage{latexsym}

\usepackage[T1]{fontenc}

\usepackage[utf8]{inputenc}

\usepackage{microtype}

\usepackage{inconsolata}

\usepackage{graphicx}

\usepackage{tabularx}    
\usepackage{amsmath,amssymb,amsfonts}
\usepackage{algorithmic}
\usepackage{multirow}
\usepackage{hyperref}
\usepackage{booktabs}
\usepackage{siunitx}
\title{GSDFuse: Capturing Cognitive Inconsistencies from Multi-Dimensional Weak Signals in Social Media Steganalysis}

\author{
  Kaibo Huang$^{1,*}$, Zipei Zhang$^{1,*}$, Yukun Wei$^{1,*}$, {\bf TianXin Zhang$^2$ ,  Zhongliang Yang$^{1,\dagger}$ , Linna Zhou$^1$}\\ 
  \thanks{Equal contribution. $^\dagger$Corresponding author.}
  $^1$Beijing University of Posts and Telecommunications,
  $^2$Beijing IntokenTech Co., Ltd., Beijing, China \\
  \texttt{\{huangkaibo, nebulazhang, weiyukun, yangzl, zhoulinna\}@bupt.edu.cn} \\
  \texttt{miki@intokentech.cn}
}

\begin{document}

\maketitle
\begin{abstract}
The ubiquity of social media platforms facilitates malicious linguistic steganography, posing significant security risks. Steganalysis is profoundly hindered by the challenge of identifying subtle cognitive inconsistencies arising from textual fragmentation and complex dialogue structures, and the difficulty in achieving robust aggregation of multi-dimensional weak signals, especially given extreme steganographic sparsity and sophisticated steganography. These core detection difficulties are compounded by significant data imbalance. This paper introduces GSDFuse, a novel method designed to systematically overcome these obstacles. GSDFuse employs a holistic approach, synergistically integrating hierarchical multi-modal feature engineering to capture diverse signals, strategic data augmentation to address sparsity, adaptive evidence fusion to intelligently aggregate weak signals, and discriminative embedding learning to enhance sensitivity to subtle inconsistencies.  Experiments on social media datasets demonstrate GSDFuse's state-of-the-art (SOTA) performance in identifying sophisticated steganography within complex dialogue environments. The source code for GSDFuse is available at \url{https://anonymous.4open.science/r/GSDFuse-B1E7}.
\end{abstract}

\section{Introduction}
\label{sec:introduction}


Steganography, the art of covert communication by embedding secret data within innocuous carriers \cite{cox2007digital}, presents a dual-use dilemma, vital for privacy, yet a potent tool for illicit activities like cyberattacks and disinformation \cite{bieniasz2018towards}. As steganographic techniques grow in sophistication, robust steganalysis becomes crucial for safeguarding digital ecosystems.

\begin{figure}[h!] 
  \centering
  \includegraphics[width=\columnwidth]{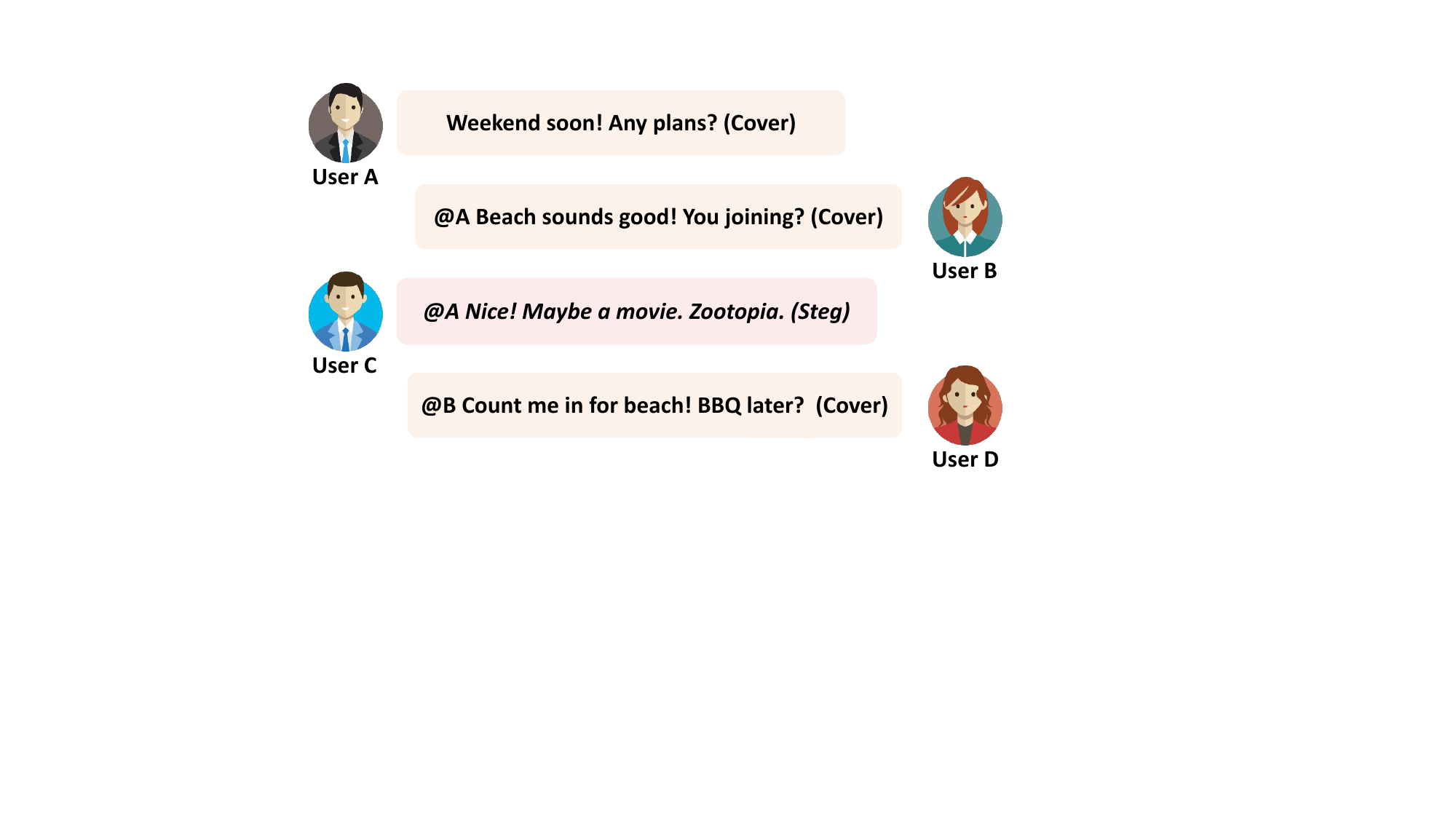} 
  \caption{An illustrative example of a social media dialogue tree. A steganographic message (Steg, User C) is subtly embedded among benign messages (Cover). This scenario highlights key challenges such as textual fragmentation, steganographic sparsity, and complex message interactions that contribute to \textit{systemic distortions} and \textit{latent multi-faceted imperceptibility} in social media steganalysis.}
  \label{fig:example_dialogue}
\end{figure}

Text's ubiquity, malleability, and persistence in information sharing establish it as a prevalent steganographic medium. The rise of social media, with its vast user base and high volume of interactive content, has further amplified text's utility as a covert communication channel \cite{Li_Wu_Lei_Wen_Ren_2018}, creating ideal environments for steganography. Modern linguistic steganography is increasingly dominated by generative approaches~\cite{yang2018rnn,yang2020vae}. These methods leverage language models to generate cover texts, embedding information by subtly altering token probabilities during generation. Notably, recent advancements have pushed towards provably secure steganography~\cite{zhang2021provably, ding2023Discop}, where the conditional probability distribution of steganographic texts becomes virtually indistinguishable from that of benign texts. Offering high embedding capacity and flexibility in cover generation, these methods also achieve great imperceptibility. 


Steganalysis, crucial for maintaining the security of public cyberspace, aims to find differences between steganographic and cover texts. Early deep learning models showed promise analyzing isolated texts~\cite{yang2019ts, wu2021linguistic, zou2020high}. However, social media's fragmented dialogues (e.g., short posts, threaded replies) severely limit extractable statistical features from single messages. This greatly hinders reliable detection based only on isolated text statistics, especially as steganography nears perfect statistical imperceptibility. This challenge drove a key shift in steganalysis: moving from breaking single-text statistical imperceptibility to using broader context to spot cognitive anomalies \cite{yang2021linguistic, 9353234}. The core idea is that isolated benign-looking messages, when viewed within larger conversational or relational structures, can reveal subtle, "unnatural" patterns. Initial progress involved integrating immediate context~\cite{yang2022linguistic}. Later work further leveraged network topology and complex feature interactions, such as by modeling connection awareness~\cite{pang2023cats} or using advanced graph architectures with attention for richer context from dialogue structures~\cite{lu2025tgca}.

Despite these advancements in leveraging context, two profound challenges remain central to robust steganalysis in social media dialogues. Firstly, modern steganography increasingly aims for \textbf{Cognitive Imperceptibility}, especially within the fragmented and interactive nature of online conversations, illustrated in Figure~\ref{fig:example_dialogue}. This means steganographic signals are designed to be indistinguishable from benign content, not just statistically at a local level, but also in terms of their naturalness and coherence within the broader dialogue flow and relational patterns. Detecting such deeply embedded content requires moving beyond surface-level features to identify subtle cues that disrupt this cognitive consistency, often only apparent when assessing the overall contextual fabric. Secondly, the difficulty of breaking this cognitive imperceptibility is compounded by the need to \textbf{Aggregate Multi-dimensional Weak Signals}. Steganography disperses faint steganographic traces across various dimensions, from deep semantic nuances and local message interactions to global dialogue structures. Effectively identifying steganography thus requires a method capable of not only constructing contextual abstracts but also of synthesizing these often sparse and individually inconclusive signals from multiple sources into a coherent judgment. Addressing these intertwined challenges is the primary motivation for our work.


The challenge of detecting steganography in social media dialogues is two-fold: while steganography achieves remarkable perceptual and statistical imperceptibility, subtle yet detectable \textbf{Cognitive Inconsistencies} can emerge within the complex conversational context. However, exploiting these inconsistencies requires overcoming the difficulty of \textbf{Aggregating Multi-dimensional Weak Signals} that indicate their presence. To address this, this paper introduces the following contributions:

\begin{itemize}
    \item We propose \textbf{GSDFuse}, a novel, multi-component collaborative method designed to exploit subtle cognitive inconsistencies and enable robust aggregation of multi-dimensional weak signals by systematically extracting and adaptively fusing diverse features to identify deeply embedded, multi-dimensional anomalies within social media dialogue trees.

    \item The proposed GSDFuse method systematically dismantles this multi-faceted challenge by synergistically integrating hierarchical feature representation from semantic node profiling to topological contextualization, adaptive cross-modal feature fusion, discriminative embedding optimization, and robust learning strategies for imbalanced and sparse data.
    \item  Experiments on large-scale, real-world social media datasets (Reddit, X [Twitter], Weibo) against mainstream steganographic algorithms (AC, HC, ADG) at various embedding rates and sparsity levels demonstrate that GSDFuse achieves SOTA performance.
\end{itemize}

\section{Related Works}


Linguistic steganalysis aims to detect steganography by identifying distinctions between steganographic and cover texts. Initial methods \cite{chen2011steganalysis, xiang2014linguistic} relied on aggregating statistical cues from n-grams or lexical features. The rise of generative steganography, producing more natural texts, spurred a shift towards deep neural networks for feature extraction from individual texts \cite{yang2019fast, zou2020high, yang2019ts, yang2020ts, peng2023text}. However, these models just struggled with statistical imperceptible, but steganography evolved to achieve statistical 
and perceptual naturalness. To address this, \citet{yang2023link} proposed leveraging external knowledge for recognizing cognitively subtle steganography, highlighting the limitations of purely internal textual features. Recognizing these limitations, a crucial paradigm shift towards context-aware steganalysis has emerged. 
Early efforts by \citet{yang2022linguistic} initiated this shift by highlighting the necessity of integrating contextual information with textual features to move beyond isolated text analysis. Subsequently, to better harness relational data, \citet{pang2023cats} focused on enhancing "connection-awareness" through graph representation learning and introduced interaction-based mechanisms for more effective fusion of textual and graph-derived features, addressing the challenge of simply concatenating disparate information. Recently, to tackle the intrinsic limitations of standard GNNs in processing complex graph-text data, \citet{lu2025tgca} incorporated Transformer and cross-attention mechanisms, aiming to refine feature aggregation from broader and more intricate social network contexts. These works represent significant steps towards more sophisticated contextual modeling in steganalysis.

Existing methods, while improving context capture, often fall short in holistically synthesizing faint, dispersed cues across diverse textual and structural dimensions, especially under extreme steganographic sparsity. This necessitates a more discriminative method capable of unveiling such elusive, context-dependent patterns, motivating GSDFuse.

\begin{figure*}[htbp]
    \centering
    \includegraphics[width=0.85\textwidth]{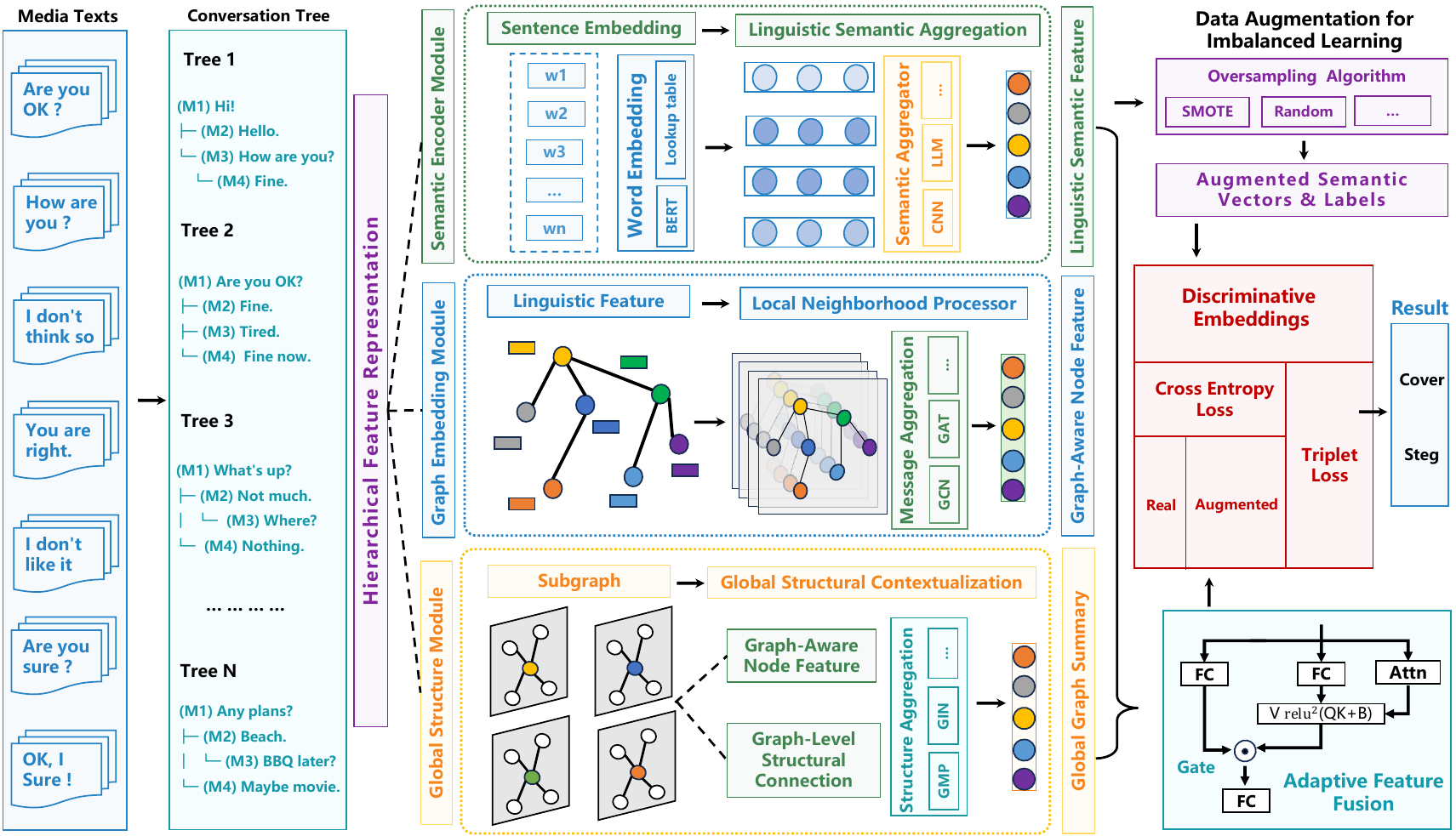}
    \caption{Our proposed GSDFuse architecture. This multi-stage pipeline is designed to identify cognitive inconsistencies by effectively aggregating multi-dimensional weak signals in social media steganalysis, leveraging diverse feature representations and optimized learning strategies.}
    \label{fig:method_gsdfuse}
\end{figure*}

\section{Preliminaries and Methodology}
\subsection{Key Challenges in Linguistic Steganalysis}
Modern linguistic steganography's primary goal is achieving both statistical and perceptual imperceptibility.
Statistical Imperceptibility, defined by \citet{Cachin2004AnIM} via KL divergence , $D_{KL}(P_C || P_S)=0$ for perfect security. Furthermore, perceptual naturalness is as critical as statistical similarity. As \citet{yang2020vae} highlighted with the "Psic Effect", statistical closeness does not guarantee perceptual naturalness, and vice versa. This inherent tension means steganographic indicators frequently manifest as dispersed, weak signals across multiple linguistic feature facets. Consequently, effective steganalysis must not only identify these cognitive inconsistencies but also achieve robust aggregation of multi-dimensional weak signals. 

\subsection{Problem Formulation and Task Definition}
We address steganalysis in social media dialogues, modeled as tree structures $\mathcal{T} = (\mathcal{V}, \mathcal{E})$. Here, $\mathcal{V} = \{v_1, \ldots, v_N\}$ is the set of $N$ message nodes, each containing textual content $x_i$, and $\mathcal{E}$ represents the reply-to relationships between them. Each node $v_i$ is associated with a binary label $y_i \in \{0, 1\}$, where $y_i=1$ indicates a steganographic text ($S$) and $y_i=0$ a benign cover text ($C$). Steganographic content is typically sparse within these dialogue trees.

The task is to develop a steganalysis model $f$ that, for a given dialogue tree $\mathcal{T}$ and all its node texts $\{x_k\}_{k=1}^N$, predicts the label $y_i$ for each node $v_i \in \mathcal{V}$. This is formulated as a node-level classification problem on the graph. For each $v_i$, the model $f$ outputs probabilities $P(Y=S | v_i, \mathcal{T})$ and $P(Y=C | v_i, \mathcal{T})$. A node $v_i$ is then classified as steganographic ($S$) if $P(Y=S | v_i, \mathcal{T}) > P(Y=C | v_i, \mathcal{T})$ .

\subsection{Overall Model Architecture}
\label{sec:overall_model_architecture}

Our proposed GSDFuse, depicted in Figure~\ref{fig:method_gsdfuse}, is a modular, multi-stage pipeline designed to systematically identify cognitive inconsistencies by effectively aggregating multi-dimensional weak signals inherent in social media steganalysis.

The \textbf{Hierarchical Feature Representation} module serves as the foundation, constructing initial node embeddings. Its primary goal is to capture both the nuanced semantic meaning of individual messages and their local relational context within the dialogue tree. This dual focus is crucial for initially exposing signals related to both perceptual naturalness and subtle structural deviations.

To address the broader, non-local nature of cognitive inconsistencies, the \textbf{Global Structural Contextualization} module complements these localized features. It derives graph-level or subgraph-level structural summaries, enabling the method to understand overarching topological patterns and anomalies that isolated node analysis would miss.

The \textbf{Adaptive Feature Fusion} module is then tasked with intelligently integrating the diverse set of features, spanning semantic content, local interactions, and global structure. Recognizing that steganographic indicators often manifest as multi-dimensional weak signals that are subtle, dispersed, and require effective aggregation, this module adaptively weighs and combines information to highlight the most salient indicators of steganography.

Finally, two critical challenges are addressed: steganographic sparsity and the need for clear class separation. The \textbf{Data Augmentation for Imbalanced Learning} module employs techniques like oversampling to mitigate the effects of sparse steganographic instances. Subsequently, the \textbf{Optimization for Discriminative Embeddings} module utilizes specialized loss functions. Its objective is to sculpt an embedding space where steganographic samples are maximally separable from benign ones, directly enhancing the model's ability to detect faint and ambiguous signals.

This multi-stage, modular architecture provides a comprehensive approach to steganalysis and offers flexibility for future enhancements and adaptations within each component to tackle evolving steganographic techniques.

\subsection{Detailed Components}
\subsubsection{Hierarchical Feature Representation} 

\paragraph{Semantic Node Profiling} 
Each message node within the dialogue tree, denoted as a sentence \(x\), is initially represented as a sequence of token identifiers \(x = (w_1, w_2, \ldots, w_L)\), where \(L\) is the length of the sentence. 
These discrete token IDs are first mapped to dense vector representations through an optimizable embedding lookup table \(E\), resulting in a sequence of token embeddings. 
Subsequently, a Semantic Composition Aggregator (SCA) module processes these token embeddings to produce a fixed-length semantic vector \(\mathbf{s}_x \in \mathbb{R}^{d_s}\) that encapsulates the core meaning of the individual message. This process can be formally expressed as:
\begin{equation}
\mathbf{s}_x  = \text{SCA}(E(w_1), E(w_2), \ldots, E(w_L)).
\label{eq:semantic_profiling}
\end{equation}
 
Various models can serve as the SCA; for example, Convolutional Neural Networks (CNNs) capture n-gram linguistic correlations, an approach demonstrated in TS-CSW~\cite{yang2020ts}.

\paragraph{Topological Contextualization} 
To incorporate structural information from the dialogue tree \(G=(V, A)\), where \(V\) denotes message nodes and \(A\) the adjacency matrix of reply-to relationships, we employ GNNs. These networks iteratively refine a node's representation by processing information from its local neighborhood. Starting with an initial node representation \(\mathbf{h}_v^{(0)} = \mathbf{s}_v\) , GNNs operate in layers. At each layer \(k\), an aggregated message \(\mathbf{m}_{\mathcal{N}(v)}^{(k)}\) is first computed from the representations of node \(v\)'s neighbors \(\mathcal{N}(v)\):
\begin{equation}
\mathbf{m}_{\mathcal{N}(v)}^{(k)} = \text{AGG}^{(k)} \left( \{ \mathbf{h}_u^{(k-1)} : u \in \mathcal{N}(v) \} \right).
\label{eq:gnn_aggregate}
\end{equation}
This message is then used to update node \(v\):
\begin{equation}
\mathbf{h}_v^{(k)} = \text{UPDATE}^{(k)} \left( \mathbf{h}_v^{(k-1)}, \mathbf{m}_{\mathcal{N}(v)}^{(k)} \right),
\label{eq:gnn_update}
\end{equation}
where \(\text{AGG}^{(k)}\) and \(\text{UPDATE}^{(k)}\) are layer-specific aggregator and update functions.

To effectively capture fine-grained structural distinctions critical for steganalysis, we can leverage powerful GNN architectures. For instance, Graph Isomorphism Network (GIN)~\cite{xu2018how} is known for its strong discriminative power, approaching the theoretical limit of the Weisfeiler-Lehman (WL) test for graph isomorphism. A GIN layer updates node representations using a sum aggregator and a Multi-Layer Perceptron (MLP):
\begin{equation}
\mathbf{a}_v^{(k)} = (1 + \epsilon^{(k)}) \cdot \mathbf{h}_v^{(k-1)} + \sum_{u \in \mathcal{N}(v)} \mathbf{h}_u^{(k-1)}.
\label{eq:gin_aggregation}
\end{equation}
This aggregated representation \(\mathbf{a}_v^{(k)}\) is then transformed by a Multi-Layer Perceptron (MLP) specific to the \(k\)-th layer:
\begin{equation}
\mathbf{h}_v^{(k)} = \text{MLP}^{(k)} \left( \mathbf{a}_v^{(k)} \right)
\label{eq:gin_mlp_update},
\end{equation}
where \(\epsilon^{(k)}\) is a learnable parameter or a fixed scalar. Through stacking such layers, the GNN produces topologically-aware node embeddings \(\mathbf{h}_v\) that reflect both semantic content and structural context within the dialogue tree.

\subsubsection{Data Augmentation for Imbalanced Learning} 
A primary challenge in steganalysis is the inherent steganographic signal sparsity, where instances of steganography (stego) are significantly outnumbered by benign cover messages. This pronounced class imbalance biases models towards the majority class, hampering their ability to detect the subtle cognitive inconsistencies introduced by advanced steganographic techniques. To address this, we employ data augmentation strategies, specifically oversampling techniques, to rebalance the training distribution and enhance the model's exposure to minority class characteristics.

One widely adopted method is the Synthetic Minority Over-sampling Technique. SMOTE  ~\cite{sun2024smote} operates by generating new synthetic minority samples in the feature space. For a given minority sample \(\mathbf{x}_i\), a new sample \(\mathbf{x}_{\text{new}}\) is created by interpolating between \(\mathbf{x}_i\) and one of its randomly selected \(k\)-nearest minority neighbors, \(\mathbf{x}_j\):
\begin{equation}
\mathbf{x}_{\text{new}} = \mathbf{x}_i + \lambda \cdot (\mathbf{x}_j - \mathbf{x}_i),
\label{eq:smote}
\end{equation}
where \(\lambda\) is a random scalar in \([0, 1]\). By applying SMOTE, typically within the learned embedding space to ensure semantic coherence of synthetic samples, we enrich the training set with diverse yet plausible stego instances. 

\begin{table*}[htbp] 
 \centering 
 \caption{F1 scores of steganalysis methods on the Sina, Tweet, and Reddit datasets using classic steganography} 
 \label{tab:HC_AC_table} 

 \resizebox{\textwidth}{!}{
\small
 \begin{tabular}{llllllllllllll}
 \toprule
 \multicolumn{2}{l}{Algorithm} & \multicolumn{6}{c}{HC} & \multicolumn{6}{c}{AC } \\ \midrule
 \multicolumn{2}{l}{Dataset} & \multicolumn{2}{c}{Sina} & \multicolumn{2}{c}{Tweet} & \multicolumn{2}{c}{Reddit} & \multicolumn{2}{c}{Sina} & \multicolumn{2}{c}{Tweet} & \multicolumn{2}{c}{Reddit} \\ 
 
\cmidrule(lr){3-4} \cmidrule(lr){5-6} \cmidrule(lr){7-8} \cmidrule(lr){9-10} \cmidrule(lr){11-12} \cmidrule(lr){13-14}
\multicolumn{2}{l}{BPW} & 1.00 & 2.59 & 1.00 & 4.35 & 2.63 & \multicolumn{1}{l}{3.42} & 2.00 & 3.27 & 5.68 & 6.94 & 0.45 & 4.69 \\ \hline
 \multirow{4}{*}{SRS=10\%} & TS-ATT$_{\text{[IWDW'21]}}$ & 65.78 & 21.39 & 68.41 & 70.89 & 84.79 & \multicolumn{1}{l}{80.93} & 36.46 & 14.86 & 64.69 & 44.02 & 88.88 & 72.88 \\
& CATS$_{\text{[ICONIP'23]}}$ & 51.07 & 15.63 & 68.44 & 60.27 & 75.13 & \multicolumn{1}{l}{68.18} & 30.96 & 21.89 & 56.09 & 49.29 & 86.77 & 62.22 \\
& TGCA$_{\text{[ICASSP'25]}}$ & 63.16 & 31.78 & 70.26 & 71.17 & 86.52 & \multicolumn{1}{l}{81.08} & 43.08 & 24.38 & 65.64 & 52.20 & 89.05 & 73.23 \\
& \textbf{ours} & \textbf{73.24} & \textbf{46.13} & \textbf{73.84} & \textbf{72.16} & \textbf{86.86} & \multicolumn{1}{l}{\textbf{82.65}} & \textbf{55.23} & \textbf{36.93} & \textbf{67.48} & \textbf{55.35} & \textbf{90.86} & \textbf{75.33} \\ \hline

\multirow{4}{*}{SRS=20\%} & TS-ATT$_{\text{[IWDW'21]}}$ & 68.24 & 47.75 & 79.21 & 77.37 & 90.25 & \multicolumn{1}{l}{84.20} & 55.17 & 24.79 & 70.34 & 61.35 & 91.89 & 76.05 \\
& CATS$_{\text{[ICONIP'23]}}$ & 65.01 & 47.91 & 80.27 & 74.25 & 86.16 & \multicolumn{1}{l}{76.67} & 53.59 & 36.06 & 66.78 & 58.51 & 91.87 & 66.80 \\
& TGCA$_{\text{[ICASSP'25]}}$ & 71.67 & 54.05 & 81.06 & 76.88 & 90.32 & \multicolumn{1}{l}{85.13} & 57.90 & 41.87 & 72.69 & 67.03 & 91.46 & 78.81 \\
& \textbf{ours} & \textbf{80.10} & \textbf{60.46} & 80.84 & \textbf{78.80} & \textbf{90.71} & \multicolumn{1}{l}{\textbf{85.30}} & \textbf{67.28} & \textbf{45.03} & 72.28 & \textbf{71.64} & \textbf{92.18} & 77.66 \\ \hline

\multirow{4}{*}{SRS=30\%} & TS-ATT$_{\text{[IWDW'21]}}$ & 75.40 & 59.07 & 86.28 & 81.49 & 90.91 & \multicolumn{1}{l}{85.72} & 55.20 & 33.28 & 77.60 & 66.45 & 91.51 & 82.45 \\
& CATS$_{\text{[ICONIP'23]}}$ & 75.95 & 59.10 & 86.41 & 78.43 & 85.98 & \multicolumn{1}{l}{79.80} & 62.09 & 48.11 & 78.38 & 68.52 & 90.61 & 78.30 \\
& TGCA$_{\text{[ICASSP'25]}}$ & 77.62 & 64.35 & 87.61 & 82.83 & 91.13 & \multicolumn{1}{l}{86.42} & 60.40 & 56.11 & 79.76 & 68.45 & 92.30 & 82.86 \\
& \textbf{ours} & \textbf{83.40} & \textbf{73.23} & \textbf{87.72} & \textbf{83.01} & \textbf{91.24} & \multicolumn{1}{l}{\textbf{86.92}} & \textbf{75.54} & \textbf{64.54} & \textbf{80.59} & \textbf{72.18} & \textbf{92.56} & \textbf{84.72} \\ \hline

\multirow{4}{*}{SRS=40\%} & TS-ATT$_{\text{[IWDW'21]}}$ & 82.89 & 58.71 & 90.36 & 87.08 & 92.64 & \multicolumn{1}{l}{88.68} & 67.33 & 40.41 & 84.93 & 71.72 & 93.18 & 84.44 \\
& CATS$_{\text{[ICONIP'23]}}$ & 83.66 & 72.84 & 90.85 & 84.19 & 89.61 & \multicolumn{1}{l}{85.35} & 68.53 & 62.79 & 82.65 & 76.37 & 93.27 & 81.38 \\
& TGCA$_{\text{[ICASSP'25]}}$ & 85.92 & 73.04 & 91.98 & 88.53 & 93.09 & \multicolumn{1}{l}{88.99} & 72.44 & 65.77 & 85.65 & 77.09 & 94.13 & 86.49 \\
& \textbf{ours} & \textbf{88.80} & \textbf{79.64} & 91.58 & 87.18 & \textbf{93.10} & \multicolumn{1}{l}{\textbf{89.43}} & \textbf{79.72} & \textbf{66.43} & \textbf{86.13} & 76.94 & \textbf{94.19} & \textbf{87.37} \\ \hline

\multirow{4}{*}{SRS=50\%} & TS-ATT$_{\text{[IWDW'21]}}$ & 84.76 & 63.97 & 93.01 & 89.43 & 92.64 & \multicolumn{1}{l}{92.07} & 69.97 & 44.08 & 87.88 & 81.04 & 95.31 & 88.15 \\
& CATS$_{\text{[ICONIP'23]}}$ & 87.66 & 79.28 & 93.39 & 88.16 & 90.77 & \multicolumn{1}{l}{88.61} & 81.03 & 73.77 & 87.51 & 80.32 & 95.22 & 85.75 \\
& TGCA$_{\text{[ICASSP'25]}}$& 88.18 & 77.79 & 94.12 & 90.45 & 93.64 & \multicolumn{1}{l}{92.51} & 77.41 & 73.27 & 88.67 & 83.75 & 96.09 & 88.98 \\
& \textbf{ours} & \textbf{91.91} & \textbf{84.36} & \textbf{94.30} & \textbf{90.85} & \textbf{94.59} & \multicolumn{1}{l}{\textbf{92.93}} & \textbf{83.98} & \textbf{76.57} & \textbf{89.23} & \textbf{85.04} & \textbf{96.17} & \textbf{90.76} \\ \bottomrule
\end{tabular}%
} 
\end{table*}

\subsubsection{Cross-Modal Feature Integration} 
Detecting steganography in social media requires effective aggregation of multi-dimensional weak signals, which are often subtly dispersed. Naive fusion methods (e.g., concatenation or averaging) for features like semantic profiles \(\mathbf{s}_v\), topological embeddings \(\mathbf{h}_v\), and global structure summaries \(\mathbf{g}_v\) often fail to resolve these nuanced and potentially conflicting indicators.

To address this, our method employs Gated Attention Unit (GAU) \cite{hua2022transformer} for \textbf{Adaptive Feature Fusion}. Given the concatenated multi-modal feature vector for a node \(v\), \(\mathbf{x}_{\text{concat},v} = [\mathbf{s}_v; \mathbf{h}_v; \mathbf{g}_v]\), where \([\cdot]\) denotes concatenation, the GAU produces an integrated feature \(\mathbf{f}_v\):
\begin{equation}
\mathbf{f}_v = \text{GAU}(\mathbf{x}_{\text{concat},v}).
\label{eq:gau_fusion}
\end{equation}

The GAU utilizes learnable gating and attention mechanisms to dynamically weigh input features and interactions per instance. This allows it to effectively arbitrate how semantic, local topological, and global structural information is combined. Adaptive fusion is crucial for the aggregation of multi-dimensional weak signals to identify subtle cognitive inconsistencies indicative of steganography, when signals are conflicting or dispersed.

\subsubsection{Optimization for Discriminative Embeddings} 
Despite adaptive fusion, feature embeddings \(\mathbf{f}_v\) of steganographic (stego) and benign messages may remain close, especially when subtle cognitive inconsistencies arise from multi-dimensional weak signals. To enhance separability, we optimize for highly discriminative embeddings using a specialized loss function.

Our approach incorporates Triplet Loss~\cite{schroff2015facenet}, a metric learning technique. It structures the embedding space by pulling same-class samples together and pushing different-class samples apart. For an anchor \(\mathbf{f}_a\) (e.g., cover), a positive \(\mathbf{f}_p\) (e.g., another cover), and a negative \(\mathbf{f}_n\) (e.g., stego), Triplet Loss aims to satisfy:
\begin{equation}
D(\mathbf{f}_a, \mathbf{f}_p) + \alpha < D(\mathbf{f}_a, \mathbf{f}_n)
\label{eq:triplet_condition},
\end{equation}
where \(D(\cdot, \cdot)\) is a distance function and \(\alpha\) is a predefined margin. The loss is:
\begin{equation}
L_{\text{triplet}} = \max(0, D(\mathbf{f}_a, \mathbf{f}_p) - D(\mathbf{f}_a, \mathbf{f}_n) + \alpha),
\label{eq:triplet_loss}
\end{equation}
where the margin \(\alpha\) compels the model to distinguish subtle steganographic traces. This focus on hard-to-separate pairs improves sensitivity. 

Triplets are selected via semi-hard negative mining. For both discriminative embeddings and accurate classification, \(L_{\text{triplet}}\) is combined with Cross-Entropy \(L_{\text{CE}}\):
\begin{equation}
L_{\text{total}} = L_{\text{CE}} + \beta \cdot L_{\text{triplet}},
\label{eq:combined_loss}
\end{equation}
where \(\beta\) balances the loss components.

\section{Experiments}
\subsection{Dataset Construction and Description}
We evaluate GSDFuse on Stego-Sandbox~\cite{yang2022linguistic}, a public social media linguistic steganalysis dataset. It offers dialogue from Twitter, Reddit, and Sina Weibo. It features steganography from three algorithms: Huffman Coding (HC), Arithmetic Coding (AC), and Adaptive Dynamic Grouping (ADG). Steganographic payload is measured by average Bits Per Word (BPW). Stego-Sandbox provides platform-specific sub-datasets, each split into training, validation, and test sets at approximately a 7:1:1 ratio. A key feature is its simulation of varying steganographic ratios via Sparsity Ratios of Stegos (SRS), ranging from 10\% to 50\%. SRS indicates the percentage of eligible texts that are steganographic.  Detailed statistics can be  Appendix~\ref{app:dataset_stats}.

\begin{table*}[htbp]
    \centering
    \caption{Additive Component Analysis for LSTSN Baseline: F1 Scores (\%) and Performance Deltas}
    \label{tab:single_component_comparison_delta_centered}
    \resizebox{\textwidth}{!}{%
    \begin{tabular}{ll*{6}{c}}
\toprule
\multicolumn{2}{l}{Algorithm} & \multicolumn{3}{c}{HC} & \multicolumn{3}{c}{AC} \\ 
 \cmidrule(lr){3-5}  \cmidrule(lr){6-8} 
\multicolumn{2}{l}{Dataset} & \multicolumn{1}{c}{Sina} & \multicolumn{1}{c}{Tweet} & \multicolumn{1}{c}{Reddit} & \multicolumn{1}{c}{Sina} & \multicolumn{1}{c}{Tweet} & \multicolumn{1}{c}{Reddit} \\ 
\multicolumn{2}{l}{BPW} & 3.23 & 4.35 & \multicolumn{1}{c}{4.17} & 4.52 & 6.94 & 5.82 \\ \hline
& LSTSN$_{\text{[TIFS'22]}}$ & 34.27 & 70.84 & \multicolumn{1}{c}{77.83} & 9.99 & 50.28 & 58.79 \\
& +GAU & 41.42 (↑7.15) & 71.36 (↑0.52) & \multicolumn{1}{c}{79.16 (↑1.33)} & 12.03 (↑2.04) & 50.77 (↑0.49) & 63.02 (↑4.23) \\
& +GIN & 35.23 (↑0.96) & 71.74 (↑0.90) & \multicolumn{1}{c}{77.74 } & 13.73 (↑3.74) & 50.11  & 60.89 (↑2.10) \\
& +Triplet Loss & 45.38 (↑11.11) & 71.29 (↑0.45) & \multicolumn{1}{c}{78.03 (↑0.20)} & 17.13 (↑7.14) & 52.84 (↑2.56) & 61.80 (↑3.01) \\
\multirow{-5}{*}{SRS=10\%} & +SMOTE & 32.65  & 70.40 & \multicolumn{1}{c}{79.13 (↑1.30)} & 10.72 (↑0.73) & 54.49 (↑4.21) & 58.12 \\ \hline
& LSTSN$_{\text{[TIFS'22]}}$ & 74.24 & 88.95 & \multicolumn{1}{c}{88.68} & 53.63 & 81.39 & 82.16 \\
& +GAU & 80.45 (↑6.21) & 90.90 (↑1.95) & \multicolumn{1}{c}{90.83 (↑2.15)} & 53.70 (↑0.07) & 83.20 (↑1.81) & 83.65 (↑1.49) \\
& +GIN & 77.67 (↑3.43) & 89.71 (↑0.76) & \multicolumn{1}{c}{89.31 (↑0.63)} & 58.59 (↑4.96) & 82.27 (↑0.88) & 83.47 (↑1.31) \\
& +Triplet Loss & 71.81  & 88.54  & \multicolumn{1}{c}{88.89 (↑0.21)} & 60.01 (↑6.38) & 81.31  & 83.07 (↑0.91) \\
\multirow{-5}{*}{SRS=50\%} & +SMOTE & 74.90 (↑0.66) & 89.41 (↑0.46) & \multicolumn{1}{c}{89.40 (↑0.72)} & 55.74 (↑2.11) & 82.71 (↑1.32) & 82.40 (↑0.24) \\ \bottomrule
    \end{tabular}%
    }
\end{table*}

\begin{figure}[htbp] 
  \centering
  \includegraphics[width=\columnwidth]{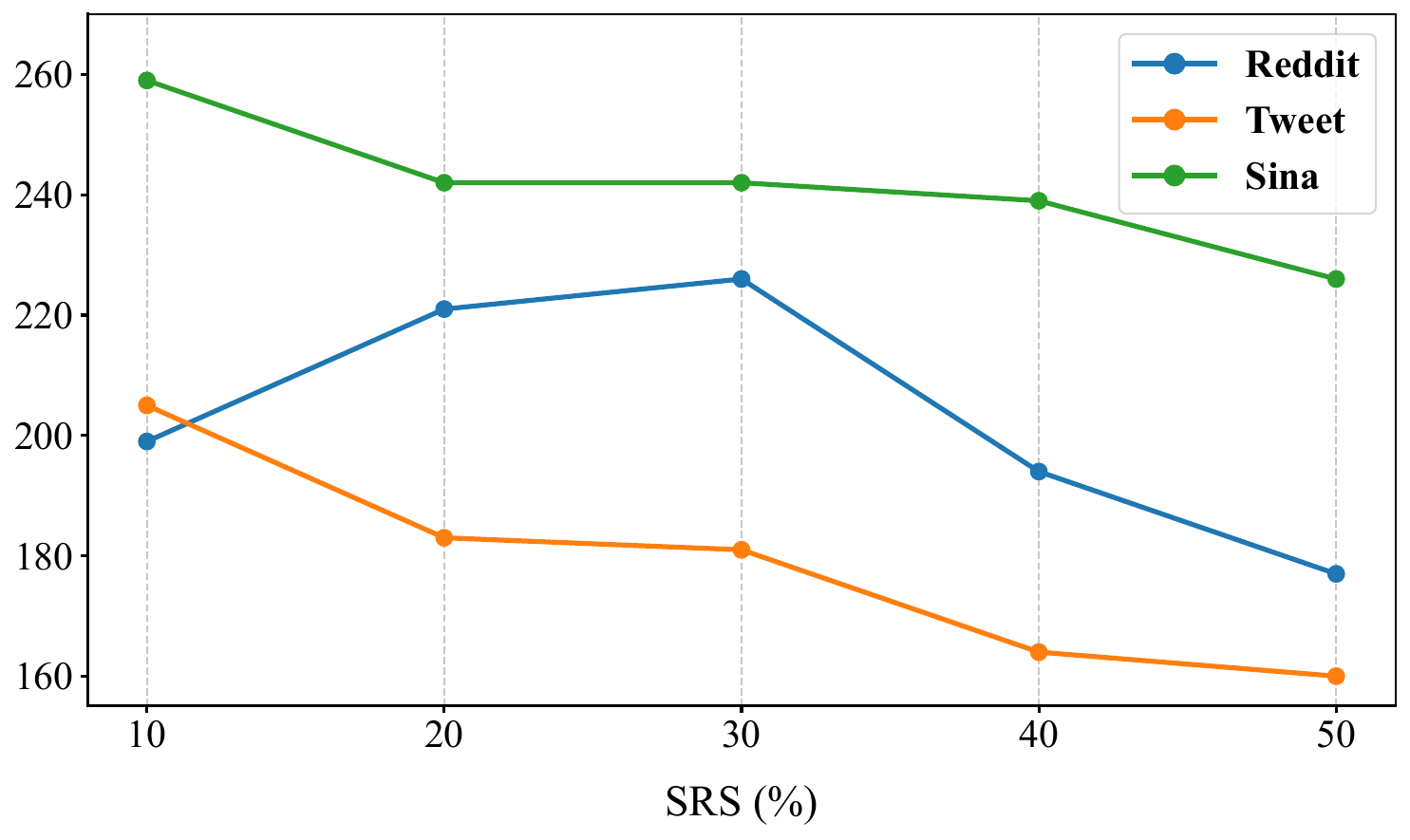} 
  \caption{Average training time (s) per epoch as a function of SRS for the Reddit, Tweet, and Sina datasets.}
  \label{fig:training_time_srs}
\end{figure}

\subsection{Experimental Setup}
Experiments ran on 10 NVIDIA L20 GPUs, taking about 2 days. Each individual model was trained for a maximum of 200 epochs. An early stopping mechanism, with a patience of 20 epochs based on the F1-score on the validation set, was employed to select the optimal model checkpoint. Figure~\ref{fig:training_time_srs} shows how average training time per epoch varies with SRS across the datasets.  We use the F1-score as our metric, ideal for imbalanced detection tasks with sparse stego data. Model performance was tested with Sparsity Ratios of Stegos (SRS) from 10\% to 50\%. Key model components include SMOTE over-sampling during training. GAU for adaptive feature fusion, GIN for global graph structure capture, and Triplet Loss as an auxiliary loss for more discriminative embeddings. Mini-batch subgraphs were sampled via a random walk strategy~\cite{graphsaint-iclr20}. The best model checkpoint was chosen by F1-score on a validation set. All reported F1-scores are averages from 3 runs. Full component-specific hyperparameter details for all components are in Appendix~\ref{app:hyperparameter_details}.

\subsection{Baseline Methods}
We compared our proposed method with four representative linguistic steganalysis baselines: TS-ATT~\cite{zou2020high}, LSTSN~\cite{yang2022linguistic}, CATS~\cite{pang2023cats}, and TGCA~\cite{lu2025tgca}. TS-ATT uses LSTM and attention for locally discordant textual features. LSTSN integrates linguistic features with social connection context; we adopted its best configuration. CATS captures graph-based social connections and employs an interaction module for deep feature integration. TGCA enhances GNN-derived topological features using Transformers to expand receptive fields and cross-attention for improved text-structure fusion.



\subsection{Experimental Results and Analysis}
\subsubsection{Steganalysis of Classic Steganography}
Table~\ref{tab:HC_AC_table} presents the F1 scores of various steganalysis methods against classic steganographic algorithms (HC and AC) across different datasets and Steganographic Ratio of Sparsity (SRS) levels. From these results, we can get the following conclusions.
Firstly, isolated text models like TS-ATT~\cite{zou2020high} primarily address statistical deviations within single messages. Their consistent underperformance highlights an inability to aggregate multi-dimensional signals from broader conversational contexts, thus failing to detect subtle cognitive inconsistencies crucial in social media dialogues.
Secondly, context-aware models such as CATS~\cite{pang2023cats} and TGCA~\cite{lu2025tgca}, while leveraging graph structures for improvement, exhibit inconsistent performance. This suggests limitations in the robust aggregation of all available multi-dimensional weak signals (textual and structural), hindering their consistent identification of nuanced cognitive inconsistencies that demand a holistic understanding of dialogue interplay.
Thirdly, our proposed GSDFuse consistently achieves superior F1 scores. Its design synergistically integrates hierarchical features with global structural contextualization, enabling effective aggregation of multi-dimensional weak signals. Furthermore, its adaptive feature fusion and discriminative embedding optimization are pivotal for unveiling subtle cognitive inconsistencies, even under high sparsity or when steganography achieves significant statistical and perceptual imperceptibility. This approach ensures robust detection of complex, context-dependent steganographic patterns.

\begin{table*}[htbp]
    \centering
    \caption{F1 scores (\%) on the Sina, Tweet, and Reddit datasets under ADG steganography.}
    \label{tab:adg_comparison}
    \renewcommand{\arraystretch}{0.7}
    \resizebox{\textwidth}{!}{%
    \tiny
    \begin{tabular}{@{}lllllllllll@{}} 
    \toprule 
    \multicolumn{1}{l}{Method} & \multicolumn{1}{l}{} & \multicolumn{3}{c}{TS-ATT$_{\text{[IWDW'21]}}$} & \multicolumn{3}{c}{LSTSN$_{\text{[TIFS'22]}}$} & \multicolumn{3}{c}{\textbf{GSDFuse}} \\
    \cmidrule(lr){3-5} \cmidrule(lr){6-8} \cmidrule(lr){9-11} 
    \multicolumn{1}{l}{Dataset} & \multicolumn{1}{l}{} & Sina    & Reddit  & \multicolumn{1}{l}{Tweet}  & Sina   & Reddit  & \multicolumn{1}{l}{Tweet}  & Sina           & Reddit         & Tweet          \\ 
    \midrule 
    \multirow{5}{*}{SRS (\%)} 
    & 10  & 2.25    & 4.70    & 7.85   & 12.61  & 8.22    & 14.07  & \textbf{13.46} & \textbf{13.85} & \textbf{17.85} \\ 
    & 20  & 2.62    & 13.74   & 16.58  & 18.62  & 20.10   & 31.76  & \textbf{20.53} & \textbf{36.41} & \textbf{33.43} \\
    & 30  & 6.21    & 22.59   & 27.81  & 22.33  & 41.82   & 45.11  & \textbf{49.13} & \textbf{47.04} & \textbf{48.22} \\
    & 40  & 9.84    & 31.39   & 41.31  & 39.09  & 51.69   & 56.29  & \textbf{60.36} & \textbf{59.47} & \textbf{58.12} \\
    & 50  & 11.14   & 41.01   & 50.85  & 43.61  & 54.52   & 67.41  & \textbf{55.80} & \textbf{67.37} & \textbf{68.31} \\
    \bottomrule 
    \end{tabular}%
    }
\end{table*}

\begin{table*}[htbp]
    \centering
   \caption{F1 scores (\%) for the ablation study of GSDFuse on the Sina, Tweet, and Reddit datasets with varying SRS levels under classic steganography.}
    \label{tab:ablation_studies}
    \resizebox{\textwidth}{!}{%
    \small
       \begin{tabular}{ll rr rr rr @{\hspace{1em}} rr rr rr}
   \toprule
    \multicolumn{2}{l}{Algorithm} & \multicolumn{6}{c}{HC} & \multicolumn{6}{c}{AC} \\
    \midrule
    \multicolumn{2}{l}{Dataset} & \multicolumn{2}{c}{Sina} & \multicolumn{2}{c}{Tweet} & \multicolumn{2}{c}{Reddit} & \multicolumn{2}{c}{Sina} & \multicolumn{2}{c}{Tweet} & \multicolumn{2}{c}{Reddit} \\
    \cmidrule(lr){3-4} \cmidrule(lr){5-6} \cmidrule(lr){7-8} \cmidrule(lr){9-10} \cmidrule(lr){11-12} \cmidrule(lr){13-14}
    \multicolumn{2}{l}{BPW} & 1.00 & 2.59 & 1.00 & 4.35 & 2.63 & 3.42 & 2.00 & 3.27 & 5.68 & 6.94 & 0.45 & 4.69 \\
    \midrule
                                   & \textbf{All Components} & \textbf{73.24} & \textbf{46.13} & \textbf{73.84} & \textbf{72.16} & \textbf{86.86} & \textbf{82.65} & \textbf{55.23} & \textbf{36.93} & \textbf{67.48} & \textbf{55.35} & \textbf{90.86} & \textbf{75.33} \\
                                   &  w/o Triplet Loss& 72.28 & 36.20 & 73.59 & 70.06 & 86.61 & 81.75  & 52.68 & 30.62 & 66.91 & 52.18 & 90.69 & 74.43 \\
                                   &  w/o SMOTE       & 68.11 & 38.83 & 72.98 & 72.04 & 86.60 & 82.23  & 47.42 & 22.54 & 66.32 & 50.69 & 90.36 & 72.89 \\
                                   &  w/o GIN         & 71.82 & 36.53 & 72.67 & 70.28 & 86.78 & 82.17  & 52.32 & 19.52 & 65.90 & 54.00 & 89.81 & 74.73 \\
    \multirow{-5}{*}{SRS=10\%}     &  w/o GAU         & 73.22 & 37.42 & 73.29 & 70.44 & 86.11 & 81.20  & 54.98 & 25.39 & 65.71 & 51.14 & 90.65 & 74.00 \\
    \midrule
                                   & \textbf{All Components} & \textbf{91.91} & \textbf{84.36} & \textbf{94.30} & \textbf{90.85} & \textbf{94.59} & \textbf{92.93} & \textbf{83.98} & \textbf{76.57} & \textbf{89.23} & \textbf{85.04} & \textbf{96.17} & \textbf{90.76} \\
                                   &  w/o Triplet Loss& 91.25 & 84.05 & 92.12 & 90.78 & 94.26 & 92.58  & 82.28 & 75.66 & 88.96 & 84.37 & 95.41 & 90.16 \\
                                   &  w/o SMOTE       & 91.66 & 84.10 & 94.26 & 90.33 & 93.69 & 92.91  & 82.30 & 67.30 & 89.15 & 83.04 & 95.91 & 89.12 \\
                                   &  w/o GIN         & 91.70 & 83.89 & 94.04 & 89.58 & 93.69 & 92.28  & 83.29 & 67.51 & 88.89 & 79.53 & 95.61 & 90.09 \\
    \multirow{-5}{*}{SRS=50\%}     &  w/o GAU         & 90.87 & 81.04 & 93.71 & 88.51 & 93.97 & 92.38  & 82.51 & 71.61 & 88.63 & 83.09 & 95.50 & 89.11 \\
    \bottomrule
    \end{tabular}
    }
\end{table*}

\subsubsection{Additive Component Analysis}
Table~\ref{tab:single_component_comparison_delta_centered} evaluates the performance impact of individually adding GSDFuse's core components to the LSTSN baseline. From these results, we can get the following conclusion.  Adding individual components such as GAU (feature fusion), GIN (structural features), and Triplet Loss (discriminative embeddings) generally improves the LSTSN baseline performance. GAU and GIN consistently enhance results, highlighting the benefits of advanced feature integration and contextual information derived from dialogue structure. Triplet Loss demonstrates particular effectiveness at lower steganographic sparsity (SRS=10\%), aiding in the separation of subtle signals. SMOTE (data augmentation) shows more varied impact but contributes positively in several scenarios. These observations affirm the individual contribution of each component designed for GSDFuse in tackling distinct facets of the steganalysis challenge.

\subsubsection{Steganalysis of Provably Secure Steganography}
From the results in Table~\ref{tab:adg_comparison}, which details performance against the provably secure ADG steganographic algorithm, we can get the following conclusion.
Detecting ADG steganography presents a significantly greater challenge for all evaluated methods, as evidenced by generally lower F1 scores compared to those achieved against classic algorithms. This is expected, given ADG's design aims for stronger security guarantees. However, despite this increased difficulty, our proposed \textbf{GSDFuse} model consistently and substantially outperforms both TS-ATT and LSTSN across all datasets (Sina, Reddit, Tweet) and at every SRS level from 10\% to 50\%. This consistent superiority, even when faced with a theoretically more robust steganographic technique, underscores GSDFuse's enhanced capability to effectively aggregate multi-dimensional weak signals and identify subtle cognitive inconsistencies. While the absolute performance is naturally impacted by ADG's sophistication, GSDFuse's robust relative performance demonstrates its advanced capacity to discern elusive steganographic traces that other methods miss.

\subsection{Ablation Studies}

Table~\ref{tab:ablation_studies} presents of GSDFuse on the Sina, Tweet, and Reddit datasets with varying SRS(10 \% and 50\%) under classic steganography. From these results, we can get the following conclusion.
Removing any single component, Triplet Loss, SMOTE, GIN, or GAU,  degrades F1 scores across all datasets compared to the complete GSDFuse model. This underscores that each module, whether for optimizing discriminative embeddings (Triplet Loss), addressing data imbalance (SMOTE), capturing structural graph features (GIN), or enabling adaptive feature fusion (GAU), contributes positively and is integral to achieving the overall robust performance of GSDFuse.

\section{Conclusion}
This paper addresses the challenges of identifying subtle cognitive inconsistencies and achieving aggregation of multi-dimensional weak signals in social media steganalysis. We proposed GSDFuse, a novel method designed to tackle these issues. GSDFuse integrates hierarchical multi-modal feature representation, employs data augmentation to counteract signal sparsity, and utilizes adaptive feature fusion to intelligently combine diverse textual and structural cues. Crucially, it optimizes for highly discriminative embeddings through a composite loss strategy, enhancing sensitivity to faint steganographic traces from both authentic and augmented data. Our extensive experiments demonstrate that GSDFuse achieves SOTA, significantly advancing the capability to detect steganography within complex conversational environments. Future work could explore extending GSDFuse to other types of covert communication or investigating its resilience against adaptive steganographic adversaries.



\section*{Limitations}
Despite GSDFuse's strong performance, its evaluation is primarily constrained by the datasets used. The Stego-Sandbox dataset, for instance, provides only token ID sequences, limiting nuanced semantic understanding by precluding raw text analysis with advanced language models. Furthermore, current public benchmarks generally lack comprehensive user-specific behavioral histories, hindering the modeling of individual communication patterns crucial for distinguishing sophisticated steganography from benign idiosyncrasies. Broader generalizability to emerging platforms, truly multimodal content, diverse cross-lingual scenarios, and low-resource environments also necessitates future evaluations on more varied and feature-rich datasets. Addressing these data-centric limitations is key for more robust steganalysis assessments.

\bibliography{custom}

\appendix 

\section{Dataset Details}
\label{app:dataset_details}
This section delineates the structure and statistical properties of the Stego-Sandbox dataset~\cite{yang2022linguistic} as employed in our research, offering a thorough account to support the main paper's findings.
\subsection{Dataset Construction and Description}
\label{app:dataset_stats}
Based on the methodology detailed in ~\citet{yang2022linguistic}, the construction of the Stego-Sandbox dataset, designed to simulate real-world social network environments for linguistic steganalysis, can be outlined through several key stages. The overall procedure involved the following steps:

Firstly, the process began with data acquisition and preprocessing. Raw textual data along with their relational information, primarily in the form of comments and replies indicating inter-text connections, were gathered from three prominent social media platforms: Twitter, Reddit, and Sina Weibo. Following collection, these raw data underwent a standard preprocessing regimen. This included essential denoising tasks such as the removal of HTTP links and hashtags, and the conversion of emojis into their corresponding textual descriptions, thereby preparing clean corpora for subsequent stages. The connections between texts, established through comments and replies, were meticulously recorded to preserve the network structure.

Secondly, the subsequent stage was dedicated to the generation of steganographic texts. To ensure the steganographic content was consistent with genuine social network scenarios, steganographic models were initially trained using the preprocessed corpora derived from each platform. Utilizing these trained models, steganographic texts were then generated by employing three state-of-the-art generative linguistic steganography algorithms: HC~\cite{yang2022linguistic}, AC~\cite{ziegler2019neural}, and ADG~\cite{zhang2021provably}. These algorithms were specifically chosen for their proficiency in minimizing the discernible statistical differences between steganographic texts and their benign counterparts. For the HC and AC algorithms, texts were also generated under a range of different embedding payloads, typically measured in BPW.

Thirdly, the next critical stage involved simulating the transmission of these steganographic texts within a modeled social network environment. This simulation was operationalized by randomly substituting a predetermined portion of the preprocessed, benign texts with the previously generated steganographic texts. A key constraint in this replacement process was that only texts which had not yet received any comments or replies were considered eligible for substitution, to reflect a scenario where the steganographic message had not yet been "delivered" or observed through interaction. Crucially, the original connection relationships of the replaced texts were preserved within the dataset's graph structure. To account for varying densities of concealed information that might be encountered in real-world networks, this substitution was performed at several distinct SRS, ranging from 10\% to 50\% in 10\% increments.

Finally, upon completion of these generation and simulation processes, the Stego-Sandbox dataset was structured and partitioned for experimental use. It was organized into three sub-datasets corresponding to Twitter, Reddit, and Sina Weibo, with each containing a specific number of total texts as detailed in the original paper (e.g., 15,639 for Twitter, 19,700 for Reddit, and 19,998 for Weibo). Each of these platform-specific sub-datasets was then randomly divided into training (comprising approximately 75\% of the data), validation (approximately 12.5\%), and testing (approximately 12.5\%) sets. An important characteristic of this partitioning was that only the connection information between texts within the designated training set was made accessible during the model training phase, thereby simulating a realistic learning environment where future interactions are unknown.

Concrete information about Stego-Sandbox can be found in Table \ref{tab:stego_sandbox_concrete_items} of the original paper~\cite{yang2022linguistic}. The authors use average hidden BPW to denote the embedding payload.

\captionsetup{font=footnotesize} 

\begin{table}[htbp]
  \centering 
  \caption{CONCRETE ITEMS ABOUT STEGO-SANDBOX} 
  \label{tab:stego_sandbox_concrete_items} 

  { 
  \small 
  \setlength{\tabcolsep}{2.5pt} 

  \noindent 
  \begin{minipage}{\columnwidth} 
  \centering 
    \textbf{A. Items about the Steganographic Text Sets.} \\
    \smallskip
    \begin{tabular}{c|ccc|ccc} 
      \hline
      \multirow{2}{*}{Steganalysis} & \multicolumn{3}{c|}{Payload (BPW)} & \multicolumn{3}{c}{Average Lengths} \\ \cline{2-7} 
                                    & Twitter     & Reddit    & Weibo    & Twitter     & Reddit     & Weibo    \\ \hline
      \multirow{5}{*}{HC}           & 1.00        & 1.00      & 1.00     & 5.88        & 10.37      & 7.81     \\
                                    & 1.90        & 1.82      & 1.86     & 7.55        & 13.78      & 8.16     \\
                                    & 2.72        & 2.63      & 2.59     & 10.13       & 15.31      & 8.40     \\
                                    & 3.57        & 3.42      & 3.23     & 12.50       & 16.75      & 8.78     \\
                                    & 4.35        & 4.17      & 3.82     & 13.77       & 17.55      & 9.03     \\ \hline
      \multirow{10}{*}{AC}          & 0.37        & 0.45      & 0.24     & 6.69        & 11.12      & 6.10     \\
                                    & 1.39        & 1.46      & 1.17     & 6.84        & 13.54      & 7.62     \\
                                    & 2.40        & 2.38      & 2.00     & 9.11        & 15.03      & 8.17     \\
                                    & 3.23        & 3.23      & 2.66     & 11.32       & 16.64      & 8.48     \\
                                    & 4.18        & 4.00      & 3.27     & 12.96       & 17.45      & 8.81     \\
                                    & 4.96        & 4.69      & 3.76     & 14.26       & 17.60      & 9.25     \\
                                    & 5.68        & 5.30      & 4.20     & 14.98       & 17.48      & 9.57     \\
                                    & 6.33        & 5.82      & 4.52     & 15.19       & 17.32      & 9.67     \\
                                    & 6.94        & 6.28      & 4.88     & 15.37       & 16.83      & 9.84     \\
                                    & 7.54        & 6.71      & 5.09     & 15.00       & 16.48      & 10.04    \\ \hline
      ADG                           & 6.93        & 8.91      & 4.13     & 12.33       & 15.91      & 9.93     \\ \hline
    \end{tabular}
  \end{minipage} 

  \vspace{0.4cm} 

  \noindent 
  \begin{minipage}{\columnwidth} 
  \centering 
    \textbf{B. Items about Processed Data} \\
    \smallskip
    \begin{tabular}{c|c|c} 
      \hline
      Platform & Number of texts & Number of connections \\ \hline
      Twitter  & 15,639          & 11,880                \\
      Reddit   & 19,700          &  9,996                \\
      Weibo    & 19,998          & 14,815                \\ \hline
    \end{tabular}
  \end{minipage} 
  } 
\end{table}



\subsection {Introduction to Steganographic Algorithms} 

\paragraph{HC}
As implemented for Stego-Sandbox, the HC steganographic method draws from approaches like RNN-Stega~\cite{yang2022linguistic} and is specifically based on Huffman coding. This technique adapts Huffman coding for textual data by conditioning on language models to embed secret information. The primary objective is to minimize statistical disruption, achieved by aligning the encoding of secret bits with the conditional probabilities of words, as detailed in~\cite{yang2022linguistic}.

\paragraph{AC}
In the context of its use for Stego-Sandbox, AC based steganography, such as described in methods like Neural Linguistic Steganography~\cite{ziegler2019neural}, encodes secret messages by mapping them to an interval based on the cumulative conditional probabilities derived from a language model. This allows for highly efficient embedding that can closely approximate the entropy of the language, thereby striving for minimal statistical detectability~\cite{yang2022linguistic}.

\paragraph{ADG}
As employed for the Stego-Sandbox dataset, ADG is a provably secure generative linguistic steganography technique~\cite{zhang2021provably}. It operates by dynamically partitioning the conditional probability distribution of the next token into several groups or 'buckets' such that the sum of probabilities in each bucket is as close to equal as possible, which has been mathematically shown to achieve a theoretical minimum in statistical difference between cover and stego texts~\cite{yang2022linguistic}.

\section{Experimental Setup and Hyperparameter Details }
\label{app:hyperparameter_details}

\subsection{Software Environment}
\label{app:hw_sw_train_time_eng}

Our experiments were conducted utilizing a hardware setup featuring ten NVIDIA L20 (48GB) GPUs. The software environment was based on Python 3.8.20, with PyTorch 2.4.1 serving as the core deep learning method and PyTorch Geometric (PyG) 2.3.1 for graph neural network functionalities. GPU computations were accelerated using CUDA 12.1. The comprehensive set of experiments involving multiple runs and configurations spanned approximately two days.

\subsection{Hyperparameter Configuration}
\label{sec:hyperparameter_configuration} 

\subsubsection{General Training Parameters}
\label{sssec:general_training_params} 
The key general parameters used throughout the training process are detailed in Table~\ref{tab:general_training_params_nonfloat}. To prevent overfitting and optimize training duration, an early stopping mechanism was also employed. Specifically, training was halted if the F1-score on the validation set did not show any improvement for 20 consecutive epochs.

\begin{center} 
    {
      \captionsetup{
        skip=3pt, 
        font=small 
      }
      \captionof{table}{General Training Parameters} 
      \label{tab:general_training_params_nonfloat} 
    } 
    \footnotesize 
    \setlength{\tabcolsep}{3pt} 
     \begin{tabular}{@{} p{0.4\columnwidth} >{\raggedleft\arraybackslash}p{0.3\columnwidth} @{}}
        \hline 
        Parameter                       & Value                                       \\ \hline
        Optimizer                       & Adam                                        \\
        Initial Learning Rate           & 0.01                                        \\
        Weight Decay                    & 0.0                                         \\
        SMOTE Batch Size                & 64                                          \\
        Total Training Epochs           & 200                                         \\
        General Dropout Rate            & 0.2                                         \\
        Random Seed                     & 42                                          \\ \hline
    \end{tabular}
\end{center}
\medskip 

\subsubsection{Model Architecture and Component Parameters}
\label{sssec:model_architecture_params_eng}

\paragraph{GNN Core}
The model's core GNN was defined with specific dimensions and operational characteristics. For feature representation, GNN intermediate layers utilized a feature dimension of 192, while sentence embeddings (when derived using the default CNN method) served as a 384-dimensional input to the GNN. The GNN architecture itself followed a `1-0-1-0' pattern, effectively comprising two GNN layers performing graph convolutions. For attention mechanisms within GAT-like layers, 8 attention heads were employed, and ReLU was used as the internal activation function.

\paragraph{Subgraph Sampling}
Subgraph sampling for training, based on the GraphSAINT methodology, was configured to use a random walk-based sampler (`rw'). This process initiated from 1000 root nodes per sampling iteration, with a random walk depth of 2. The target node budget per subgraph, dictating the approximate subgraph size, was set to 2000, and a sample coverage of 50 was maintained.

\paragraph{SMOTE}
To address class imbalance, SMOTE could be utilized. When active, SMOTE was configured with 5 k-nearest neighbors and a random state seed of 42. It generated 64 synthetic samples per mini-batch, and the loss contribution from these SMOTE-generated samples was weighted by a factor of 0.5.

\paragraph{GAU} 
GAU included setting the internal query and key dimension ratio to input feature dimension at 1/4 and an MLP expansion factor of 2. Laplacian Attention was activated, and other GAU hyperparameters (e.g., number of attention heads, dropout) utilized default values from the \texttt{flash\_pytorch.GAU} library.

\paragraph{GIN} 
GIN was configured with 2 GIN convolution layers, an internal dropout rate of 0.1 per layer, an enabled learnable epsilon parameter, and used sum aggregation.

\paragraph{Triplet Loss} 
\begin{sloppypar} 
For experiments employing Triplet Loss, the margin $\alpha$ was set to $1.0$, using Euclidean distance ($p=2$). The default sample mining strategy involved no specific hard sample mining or a semi-hard strategy, though hard negative/positive mining could be enabled via command-line arguments (--use\_hard\_mining=True and --mining\_strategy=`hard'). The weighting factor for the Triplet Loss component in the total loss computation $\lambda_{\text{triplet}}$ was $0.1$.
\end{sloppypar}
\subsection{Other Implementation Details}

\subsubsection{Loss Function Composition}
\label{sssec:loss_function_composition} 
The total loss function, $L_{\text{total}}$, is formulated as a weighted sum of three distinct components.The precise composition is given by the equation below:
\[ L_{\text{total}} = L_{\text{CE}} + \lambda_{\text{SMOTE}} \cdot L_{\text{SMOTE\_CE}} + \lambda_{\text{triplet}} \cdot L_{\text{triplet}} \]
In this formulation, $L_{\text{CE}}$ represents the standard classification loss, which is typically the Cross-Entropy loss. The term $L_{\text{SMOTE\_CE}}$ denotes the classification loss computed specifically on synthetic samples generated by the SMOTE, and this component is weighted by $\lambda_{\text{SMOTE}}$, set to 0.5. Finally, $L_{\text{triplet}}$ signifies the Triplet Loss, incorporated to enhance the learning of discriminative feature embeddings, with its corresponding weighting factor $\lambda_{\text{triplet}}$ set to 0.1.

\end{document}